\documentclass[fleqn]{2023SCGE}
\usepackage{orcidlink}
\usepackage{multirow}

\newcommand{\nc}{\newcommand*}
\nc{\mU}{{\mathcal{U}}}
\nc{\Msun}{M_\odot}   
\nc{\red}[1]{\textcolor{red}{#1}}
\nc{\Eq}[1]{Eq.~\eqref{#1}}     
\nc{\Fig}[1]{Fig.~\ref{#1}}     
\nc{\Table}[1]{Table~\ref{#1}}  
\nc{\Sec}[1]{Sec.~\ref{#1}}     
\newcommand{\be}{\begin{equation}}
\newcommand{\ee}{\end{equation}}

\def\({\left(}
\def\){\right)}
\def\[{\left[}
\def\]{\right]}

\begin{document}
\ensubject{subject}

\ArticleType{Article}
\SpecialTopic{SPECIAL TOPIC: }
\Year{2025}
\Month{}
\Vol{}
\No{}
\DOI{}
\ArtNo{}
\ReceiveDate{}
\AcceptDate{}
\OnlineDate{}

\title{Constraining the nonstandard propagating gravitational waves in the cosmological background with GWTC-3}{Constraining the nonstandard propagating gravitational waves in the cosmological background with GWTC-3} 

\author[1,2]{Zu-Cheng~Chen\orcidlink{0000-0001-7016-9934}}{}
\author[3]{Lang~Liu\orcidlink{0000-0002-0297-9633}}{liulang@bnu.edu.cn}

\AuthorMark{Z.-C. Chen}
\AuthorCitation{Z.-C. Chen, L. Liu}

\address[1]{Department of Physics and Synergetic Innovation Center for Quantum Effects and Applications, Hunan Normal University, Changsha, Hunan 410081, China}
\address[2]{Institute of Interdisciplinary Studies, Hunan Normal University, Changsha, Hunan 410081, China}
\address[3]{Faculty of Arts and Sciences, Beijing Normal University, Zhuhai 519087, China}

\abstract{
The detection of gravitational waves (GWs) has opened a new window to test the fundamental nature of gravity. We present constraints on the nonstandard propagation of GWs using the spectral siren method applied to binary black hole (BBH) mergers from the third Gravitational-Wave Transient Catalog (GWTC-3). The spectral siren method exploits the redshift distribution of BBHs to probe the cosmic expansion history and break degeneracies between cosmology and modified gravity effects. We focus on the friction term $\nu$ in the nonstandard GW propagation equation, which characterizes the running of the Planck mass. Assuming the standard $\Lambda$CDM cosmology, we find $\nu = -1.1^{+3.9}_{-1.1}$, $\nu = 0.5^{+3.5}_{-2.6}$, and $\nu = 0.7^{+3.1}_{-2.3}$ (median and $90\%$ credible interval) for the Truncated, Power Law + Peak, and Broken Power Law mass models, respectively. These results improve upon previous constraints from the bright siren event GW170817 by an order of magnitude, owing to the higher redshifts of BBHs in GWTC-3, which reach up to $z \sim 1$. Our result suggests that the propagation of GWs is consistent with the predictions of general relativity, placing limits on modified gravity theories that predict a time-varying Planck mass. As the sensitivity of GW detectors improves, the spectral siren method will provide a powerful tool for testing gravity on cosmological scales and probing the physics of the early Universe.
}

\maketitle

\begin{multicols}{2}
\section{Introduction}\label{section1}
The quest to understand the fundamental nature of gravity and its role in the Universe has led to the exploration of modified gravity theories. Two primary motivations drive the pursuit of these alternative theories: the need to explain cosmic acceleration and the desire to find a substitute for dark matter. The observed accelerated expansion of the Universe \cite{Riess:1998cb,Perlmutter:1998np} and the presence of dark matter \cite{Bertone:2004pz,Clowe:2006eq} have challenged the standard model of cosmology based on general relativity \Authorfootnote (GR) and the $\Lambda$CDM paradigm. Modified gravity theories  aim to address these challenges by introducing additional degrees of freedom or modifying the gravitational action \cite{Nojiri:2010wj,Clifton:2011jh,Joyce:2014kja,Koyama:2015vza,Nojiri:2017ncd}.  These modifications can manifest in various ways, including altering the propagation of gravitational waves (GWs) through the cosmological background \cite{Saltas:2014dha,Nishizawa:2017nef,Arai:2017hxj}.

Modified gravity theories typically introduce changes in the weak-field regime at large scales, while reducing to GR in the strong-field regime through screening mechanisms such as Chameleon \cite{Khoury:2003aq,Khoury:2003rn}, Vainshtein \cite{Vainshtein:1972sx,Babichev:2013usa}, or Symmetron \cite{Hinterbichler:2010es,Hinterbichler:2011ca}. These screening mechanisms ensure that the theories remain consistent with local tests of gravity \cite{Will:2014kxa,Burrage:2017qrf}. Cosmological tests of modified gravity often focus on the propagation of GWs rather than their generation \cite{Saltas:2014dha,Nishizawa:2017nef,Arai:2017hxj}. Even if the modification to gravity is a tiny effect, the propagation of GWs over long distances can accumulate the effect, making it detectable by current and future GW detectors \cite{Belgacem:2017ihm,Belgacem:2018lbp}.

The detection of GWs by the Advanced LIGO and Advanced Virgo detectors \cite{LIGOScientific:2016aoc,LIGOScientific:2017vwq,LIGOScientific:2018mvr,LIGOScientific:2020ibl,KAGRA:2021vkt} has opened up a new observational window to test modified gravity theories, complementing the existing probes based on electromagnetic waves. One powerful method to constrain these theories is the bright siren approach, which relies on the detection of GWs with associated electromagnetic counterparts \cite{Schutz:1986gp,Holz:2005df,Dalal:2006qt,Nissanke:2009kt}. A notable example of a bright siren is the binary neutron star merger event GW170817 with its accompanying kilonova emission \cite{LIGOScientific:2017vwq,LIGOScientific:2017ync}. However, the applicability of the bright siren method is limited by the rarity of such events and the need for precise sky localization.

An alternative approach is the ``spectral siren" method, which exploits features in the mass distribution of compact binaries, such as gaps or peaks, to constrain their redshift distribution \cite{Ezquiaga:2022zkx}. By tracking the redshifting of these features, one can probe the expansion history independently of electromagnetic observations. While spectral sirens may not measure the Hubble constant as precisely as bright sirens~\cite{LIGOScientific:2021aug}, they have the potential to significantly outperform them in constraining departures from GR in GW propagation, as we will demonstrate in this paper.

In this paper, we focus on how nonstandard propagation of GWs can be constrained using the spectral siren method. Modified gravity theories often predict a departure from the standard GR wave equation, introducing additional terms that can be interpreted as a friction effect \cite{Saltas:2014dha,Nishizawa:2017nef,Ezquiaga:2017ekz}. This modification accumulates over the vast distances traveled by cosmological GWs, making their propagation a sensitive probe of even small deviations from GR \cite{Nishizawa:2017nef,Arai:2017hxj,Lagos:2019kds}. We aim to constrain these nonstandard propagation effects using the third GW Transient Catalog (GWTC-3) \cite{KAGRA:2021vkt}, leveraging the spectral siren method to break degeneracies between cosmology and modified gravity. Remarkably, we find that spectral sirens can constrain the friction term in modified GW propagation an order of magnitude better than bright sirens, highlighting their complementary role in probing gravity.


\section{Theoretical Framework for Nonstandard GW Propagation}\label{framework}
The propagation of GWs in the cosmological background can be modified by alternative theories of gravity, leading to observable deviations from the predictions of GR. A generalized framework has been developed to describe and test such nonstandard GW propagation, based on the effective field theory approach~\cite{Nishizawa:2017nef,Arai:2017hxj,Nishizawa:2019rra,Zhu:2023wci,Lin:2024pkr}.

In this framework, the propagation equation for GWs is given by
\begin{equation}
h_{ij}^{\prime\prime} + (2 + \nu) \mathcal{H} h_{ij}^\prime + (c_\mathrm{T}^2 k^2 + a^2\mu^2)h_{ij} = a^2\Gamma\gamma_{ij},
\end{equation}
where $h_{ij}$ is the metric perturbation, $\mathcal{H} \equiv a^\prime/a$ is the conformal Hubble parameter, and the prime denotes the derivative with respect to conformal time. The equation contains four time-dependent parameters $(\nu, c_\mathrm{T}, \mu, \Gamma)$ that characterize possible deviations from GR:

\begin{itemize}
\item $\nu$ is related to the time variation of the gravitational coupling strength, arising from the running of the Planck mass~\cite{Bellini:2014fua}.
\item $c_\mathrm{T}$ is the propagation speed of GWs, which can differ from the speed of light.
\item $\mu$ represents the effective mass of the graviton.
\item $\Gamma\gamma_{ij}$ is a source term due to anisotropic stress, which can modify the GW amplitude.
\end{itemize}

In the limit where $\nu = 0$, $c_\mathrm{T} = 1$, $\mu = 0$, and $\Gamma = 0$, the propagation equation reduces to the standard GR case:
\begin{equation}
h_{ij}^{\prime\prime} + 2\mathcal{H} h_{ij}^\prime + k^2 h_{ij} = 0.
\end{equation}
This shows that GR is a special case within the generalized framework, corresponding to the absence of any modification parameters.

In this paper, we focus on the effects of the $\nu$ parameter, which is the least constrained by current observations. The GW170817 event has placed a stringent bound on the speed of GWs, $|c_\mathrm{T} - 1| \lesssim 10^{-15}$ \cite{LIGOScientific:2017zic}, allowing us to set $c_\mathrm{T} = 1$. The graviton mass $\mu$ is also tightly constrained by various observations, such as solar system tests and binary pulsar observations \cite{deRham:2016nuf}, justifying the assumption $\mu = 0$. The source term $\Gamma$ is neglected for simplicity, as it is expected to be subdominant in most scenarios.
Under these assumptions, the propagation equation simplifies to
\begin{equation}
h_{ij}^{\prime\prime} + (2 + \nu) \mathcal{H} h_{ij}^\prime + k^2 h_{ij} = 0.
\end{equation}
This equation describes the nonstandard propagation of GWs due to the running of the Planck mass, characterized by the $\nu$ parameter \cite{Nishizawa:2017nef}.

Assuming $\nu$ varies slowly over cosmological timescales, while the GW wavelength is much smaller than the horizon scale, the WKB solution to the propagation equation can be expressed as \cite{Nishizawa:2017nef}
\begin{equation}
h = C_\mathrm{MG} h_\mathrm{GR},
\end{equation}
where $h_\mathrm{GR}$ is the GR waveform and the correction factor $C_\mathrm{MG}$ is given by
\begin{equation}
C_\mathrm{MG} = e^{-\mathcal{D}},
\end{equation}
with $\mathcal{D}$ quantifying the modification to the GW amplitude:
\begin{equation}
\mathcal{D} = \frac{1}{2} \int_0^z \frac{\nu}{1+z^\prime} \mathrm{d}z^\prime.
\end{equation}

The luminosity distance in the presence of nonstandard GW propagation can be derived by considering the energy flux of GWs. In GR, the luminosity distance $d_L^\mathrm{GR}$ is related to the redshift $z$ by \cite{Maggiore:2007ulw}
\begin{equation}
d_L^\mathrm{GR}(z) = (1+z) \int_0^z \frac{c\, \mathrm{d}z^\prime}{H(z^\prime)},
\end{equation}
where $c$ is the speed of light and $H(z)$ is the Hubble parameter as a function of redshift.

In the nonstandard propagation scenario, the GW amplitude is modified by the factor $C_\mathrm{MG}$, which leads to a modified luminosity distance $d_L^\mathrm{MG}$ \cite{Belgacem:2018lbp, Belgacem:2019pkk}:
\begin{equation}\label{dLMG}
d_L^\mathrm{MG}(z) = \frac{d_L^\mathrm{GR}(z)}{C_\mathrm{MG}(z)} = d_L^\mathrm{GR}(z) e^{\mathcal{D}(z)}.
\end{equation}
For the purposes of this paper, we assume that the friction term $\nu$ remains constant within the observable redshift range (i.e., for $z \lesssim 2$), following the approach of~\cite{Amendola:2017ovw}.
This parameterization differs from recent works that have applied the spectral siren method to the GWTC-3 catalog to constrain various other parametrized frameworks of modified gravity, such as the $\Xi_0$ or the $c_M$ parameters, while employing different mass models~
\cite{Mancarella:2021ecn,Leyde:2022orh}.
Combining the expressions for $d_L^\mathrm{GR}(z)$ and $\mathcal{D}(z)$, we obtain the relation between the modified luminosity distance and redshift~\cite{Nishizawa:2017nef,Tian:2019vkc}:
\begin{equation}
d_L^\mathrm{MG}(z) = (1+z)^{\nu/2} d_{\mathrm{GR}}
\end{equation}
where 
\begin{equation}
d_{\mathrm{GR}}  =  \frac{(1+z)}{H_0} \int_{0}^{z} \frac{dz'}{\sqrt{\Omega_\mathrm{m} (1+z')^3 + (1-\Omega_\mathrm{m})}}.
\end{equation}
Here, $H_0$ denotes the Hubble constant and $\Omega_\mathrm{m}$ is the matter density parameter at present. Besides, $(1-\Omega_\mathrm{m})$ represents the fraction of the Universe's energy density in the form of dark energy, assuming a flat Universe and neglecting the contribution of radiation.

\Eq{dLMG} shows that the nonstandard propagation of GWs, characterized by the $\nu$ parameter, leads to a modification of the luminosity distance-redshift relation compared to the GR case. By measuring the luminosity distances of GW events at different redshifts, it is possible to constrain the $\nu$ parameter and test the validity of GR on cosmological scales \cite{Belgacem:2018lbp, Belgacem:2019pkk, Mukherjee:2020mha, Mastrogiovanni:2020mvm}.

This simplified framework allows us to isolate and constrain the effects of the $\nu$ parameter on the propagation of GWs using the GWTC-3 catalog from the LIGO-Virgo-KAGRA Collaboration \cite{KAGRA:2021vkt}. By comparing the observed GW signals with the predictions of this nonstandard propagation model, we can place bounds on the possible time variation of the gravitational coupling strength, providing a test of GR on cosmological scales.

\section{Methodology}\label{methodology}
We utilize a set of 42 binary black holes (BBHs) from the GWTC-3 catalog \cite{KAGRA:2021vkt}, selected based on a network-matched filter signal-to-noise ratio threshold of 11 and an inverse false alarm rate higher than 4 years, as outlined in \cite{LIGOScientific:2021aug,Chen:2022fda}. Our analyses incorporate combined posterior samples derived from the IMRPhenom \cite{Thompson:2020nei,Pratten:2020ceb} and SEOBNR \cite{Ossokine:2020kjp,Matas:2020wab} waveform families.

GW experiments measure the luminosity distance $D_{\mathrm{L}}$ and redshifted masses $m_{1}^{\mathrm{det}}, m_{2}^{\mathrm{det}}$ for each BBH event, rather than the source redshift $z$ and intrinsic masses $m_{1}$, $m_{2}$. These quantities are connected via the relation:
\begin{equation}\label{mz}
m_{i}=\frac{m_{i}^{\text{det}}}{1+z(D_\mathrm{L}^\mathrm{MG} ; H_{0}, \Omega_{\mathrm{m}})}.
\end{equation}
By characterizing the source mass distribution, this relation enables probing the cosmic expansion history without relying on redshift information from electromagnetic counterparts \cite{Taylor:2011fs,Taylor:2012db}.

Given the observed data $\textbf{d} = \{d_1, d_2, \cdots, d_{N_{\mathrm{obs}}}\}$ from $N_{\mathrm{obs}}$ BBH merger events, the total number of events is modeled as an inhomogeneous Poisson process, yielding the likelihood \cite{Loredo:2004nn,Thrane:2018qnx,Mandel:2018mve,Mastrogiovanni:2023zbw}:
\begin{equation}
\label{eq:L1}
\mathcal{L}(\textbf{d}|\Lambda) \propto e^{-N_{\mathrm{exp}}} \prod_{i=1}^{N_{\mathrm{obs}}} \int T_{\mathrm{obs}}\, \mathcal{L} (d_{i}| \theta)\, \mathcal{R}_{\mathrm{pop}}(\theta|\Lambda) d \theta,
\end{equation}
where $N_{\exp}(\Lambda) \equiv \xi(\Lambda) T_{\mathrm{obs}}$ is the expected number of detections over the observation timespan $T_{\mathrm{obs}}$, and $\mathcal{L} (d_{i}|\theta)$ represents the likelihood for the $i$th GW event, obtained by reweighing the individual event posterior with the prior on $\theta$. The term $\xi(\Lambda)$ accounts for selection biases for a population with parameters $\Lambda$:
\begin{equation}
\label{eq:xi}
\xi(\Lambda) = \int P_{\mathrm{det}}(\theta)\, \mathcal{R}_{\mathrm{pop}}(\theta|\Lambda)\, \mathrm{d} \theta,
\end{equation}
where $P_{\mathrm{det}}(\theta)$ is the detection probability as a function of source parameters $\theta$ \cite{OShaughnessy:2009szr}.

\begin{table*}[tbp!]
	\centering
    \caption{\label{tab:priorss} Parameters and their prior distributions used in the Bayesian parameter estimations.}
		\begin{tabular}{lll}
			\hline\hline
			\textbf{Parameter} & \textbf{Description} & \textbf{Prior} \\
			\hline
            $\nu$ & friction parameter. & $\mU(-5, 6)$\\
			$R_0$ & PBH merger rate today in $\mathrm{Gpc}^{-3} \mathrm{yr}^{-1}$. & $\mU(0, 200)$\\
			$H_0\,[\mathrm{km}\, \mathrm{s}^{-1} \mathrm{Mpc}^{-1}]$ & Hubble constant. & $\mU(20, 120)$ \\
			$\Omega_m$ & Present-day matter density of the Universe. & $\mU(0, 1)$\\	
			$\gamma$ & Slope of the power law regime for the rate evolution before the point $z_p$. & $\mU(0, 12)$\\
			$\kappa$ & Slope of the power law regime for the rate evolution after the point $z_p$. & $\mU(0, 6)$\\
			$z_p$ & Redshift turning point between the powerlaw regimes with $\gamma$ and $\kappa$. & $\mU(0, 4)$\\   	
   \hline
   \multicolumn{3}{c}{Truncated} \\[1pt]
   $\alpha$ & Spectral index for the power law of the primary mass distribution. & $\mU(2, 8)$\\
			$\beta$ & Spectral index for the power law of the mass ratio distribution. & $\mU(-3, 7)$\\
			$m_{\min}\,[\Msun]$ & Minimum mass of the power law component of the primary mass distribution. & $\mU(2, 7)$\\
			$m_{\max}\,[\Msun]$ & Maximum mass of the power law component of the primary mass distribution. & $\mU(50, 200)$\\
   \hline
   \multicolumn{3}{c}{Power Law + Peak} \\[1pt]
			$\alpha$ & Spectral index for the power law of the primary mass distribution. & $\mU(2, 8)$\\
			$\beta$ & Spectral index for the power law of the mass ratio distribution. & $\mU(-3, 7)$\\
			$m_{\min}\,[\Msun]$ & Minimum mass of the power law component of the primary mass distribution. & $\mU(2, 7)$\\
			$m_{\max}\,[\Msun]$ & Maximum mass of the power law component of the primary mass distribution. & $\mU(50, 200)$\\
			$\lambda_g$ & Fraction of the model in the Gaussian component. & $\mU(0, 0.2)$\\
			$\mu_g\,[\Msun]$ & Mean of the Gaussian component in the primary mass distribution. & $\mU(20, 50)$\\
			$\sigma_g\,[\Msun]$ & Width of the Gaussian component in the primary mass distribution. & $\mU(0.4, 10)$\\
			$\delta_m\,[\Msun]$ & Range of mass tapering at the lower end of the mass distribution. & $\mU(0, 10)$\\
   \hline
   \multicolumn{3}{c}{Broken Power Law} \\[1pt]
   $\alpha_1$ & Power law slope of the primary mass distribution for masses below $m_\mathrm{break}$ & $\mU(2, 8)$\\
	$\alpha_2$ & Power law slope of the primary mass distribution for masses above $m_\mathrm{break}$ & $\mU(2, 8)$\\
 
			$\beta$ & Spectral index for the power law of the mass ratio distribution. & $\mU(-3, 7)$\\
			$m_{\min}\,[\Msun]$ & Minimum mass of the power law component of the primary mass distribution. & $\mU(2, 7)$\\
			$m_{\max}\,[\Msun]$ & Maximum mass of the power law component of the primary mass distribution. & $\mU(50, 200)$\\			
			$b$ & The fraction of the way between $m_{\min}$ and $m_{\max}$ at which the primary mass distribution breaks. & $\mU(0, 1)$\\
			$\delta_m\,[\Msun]$ & Range of mass tapering at the lower end of the mass distribution. & $\mU(0, 10)$\\
			\hline
		\end{tabular}
\end{table*}

In practice, $\xi(\Lambda)$ is estimated using simulated injections \cite{KAGRA:2021vkt} and approximated through a Monte Carlo integral over found injections \cite{KAGRA:2021duu}:
\begin{equation}
\xi(\Lambda) \approx \frac{1}{N_{\mathrm{inj}}} \sum_{j=1}^{N_{\text{found}}} \frac{\mathcal{R}_{\mathrm{pop}}(\theta_j | \Lambda)}{p_{\mathrm{draw}}(\theta_j)},
\end{equation}
where $N_{\text{inj}}$ is the total number of injections, $N_{\text{found}}$ is the count of successfully detected injections, and $p_{\mathrm{draw}}$ is the probability density function used to draw the injections. Using the posterior samples from each event, the hyper-likelihood in Eq.~\eqref{eq:L1} is computed as:
\begin{equation}
\label{eq:L2}
\mathcal{L}(\textbf{d}|\Lambda) \propto e^{-N_{\mathrm{exp}}} \prod_{i=1}^{N_{\mathrm{obs}}} T_{\mathrm{obs}} \left\langle \frac{\mathcal{R}_{\mathrm{pop}}(\theta|\Lambda)}{\pi(\theta)} \right\rangle,
\end{equation}
where $\langle\cdots\rangle$ denotes the weighted average over posterior samples of $\theta$, and $\pi(\theta)$ refers to the priors on source parameters used to construct the individual event posteriors.

The merger rate $\mathcal{R}_{\mathrm{pop}}$ in Eq.~\eqref{eq:L1} is given by:
\begin{equation}
\label{Rpop}
\mathcal{R}_{\mathrm{pop}} = \frac{1}{1+z} \frac{dV_\mathrm{c}}{dz} R_{0}\, \pi(z)\, \pi(m_1)\, \pi(m_2| m_1),
\end{equation}
where $R_0$ is the local merger rate, $dV_\mathrm{c}/dz$ is the differential comoving volume, and $1/(1 + z)$ accounts for time dilation between the source and detector frames \cite{Fishbach:2018edt}.
The redshift distribution $\pi(z)$ is modeled as a double power law \cite{Fishbach:2018edt,Callister:2020arv}:
\begin{equation}
\label{eq:pz}
\pi(z| \gamma, \kappa, z_{\mathrm{p}}) = \left[1+\left(1+z_{\mathrm{p}}\right)^{-\gamma-\kappa}\right] \frac{(1+z)^\gamma}{1+\left[(1+z) / (1+z_\mathrm{p})\right]^{\gamma + \kappa}},
\end{equation}
with slopes $\gamma$ and $\kappa$, and a characteristic redshift $z_{\mathrm{p}}$. This parameterization is motivated by the potential correlation between the binary formation rate and the star formation rate \cite{Madau:2014bja}.

We consider three different mass distributions, which are described in more detail in Ref.~\cite{KAGRA:2021duu,LIGOScientific:2021aug}.
\begin{itemize}
    \item \textbf{Truncated model}: The truncated mass model is characterized by a power law distribution for the primary mass $m_1$ with spectral index $\alpha$ and a sharp cut-off at $m_{\min}$ and $m_{\max}$. The mass ratio $q \equiv m_2/m_1$ follows a power law distribution with spectral index $\beta$.
    \item \textbf{Power Law + Peak model}: The primary mass distribution $\pi(m_1)$ follows a ``Power Law + Peak" model~\cite{KAGRA:2021duu}, combining a truncated power law $\mathcal{P}$ with slope $\alpha$ and a Gaussian component $\mathcal{G}$ with mean $\mu_{\mathrm{g}}$ and standard deviation $\sigma_{\mathrm{g}}$. The mixing fraction between the two components is governed by $\lambda_{\mathrm{g}}$. A sigmoid-like window function $S$ ensures a smooth transition at the lower mass cutoff $m_{\min}$ over a scale $\delta_m$.
The secondary mass distribution $\pi(m_2)$ is modeled as a truncated power law with slope $\beta$ between $m_{\min}$ and the primary mass $m_1$ \cite{KAGRA:2021duu}.
    \item \textbf{Broken Power Law model}: The Broken Power Law model describes the distribution of the primary mass $m_1$ as a power law between a minimum mass $m_{\min}$ and a maximum mass $m_{\max}$. The model is characterized by two power law slopes, $\alpha_1$ and $\alpha_2$, and a breaking point at $m_{\text{break}} = b\, (m_{\max} - m_{\min})$, where $b \in [0, 1]$.
\end{itemize}

This comprehensive modeling approach allows for a detailed characterization of the merger rate, redshift distribution, and mass distributions of BBH systems, enabling a thorough investigation of their properties and potential origins. The inclusion of the redshift distribution and its potential link to the star formation rate provides insights into the formation channels and evolutionary history of these systems \cite{Fishbach:2018edt,Callister:2020arv}. The mass distributions, with the power law and Gaussian components, capture the observed features in the black hole mass spectrum and aid in understanding the formation mechanisms and possible existence of a mass gap \cite{KAGRA:2021duu}.

\section{Results and Discussion}\label{results}
\begin{figure}[H]
	\centering
	\includegraphics[width=0.45\textwidth]{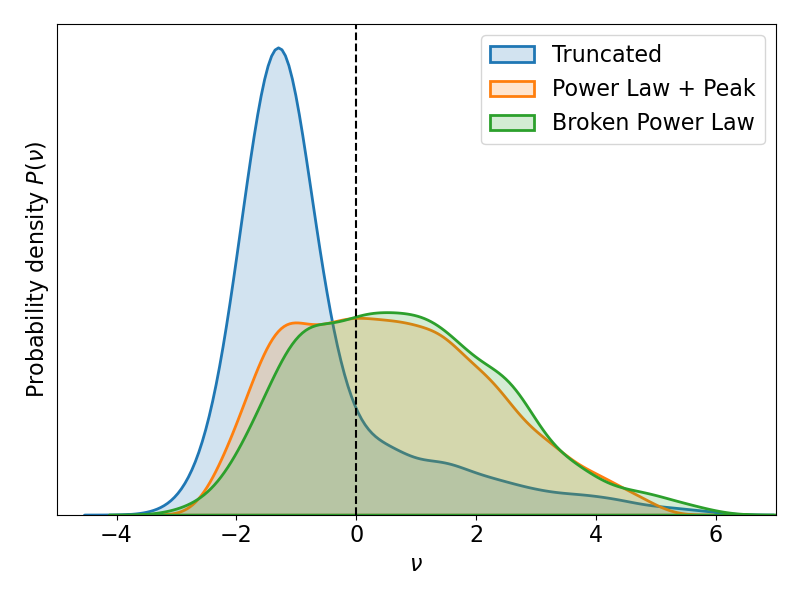}
	\caption{\label{post_nu}Marginalized posterior distribution for the $\nu$ parameter constrained by the GWTC-3 catalog for Truncated, Power Law+Peak, and Broken Power Law mass models, respectively. The black dashed line indicates the case where $\nu=0$, corresponding to the prediction from GR.}
\end{figure}

The prior distributions for the model parameters are summarized in \Table{tab:priorss}. 
The likelihood function is evaluated using the \texttt{ICAROGW} package \cite{Mastrogiovanni:2023zbw}. The parameter space is explored using the \texttt{dynesty} nested sampling algorithm \cite{Speagle:2019ivv}, which is called from the \texttt{Bilby} package \cite{Ashton:2018jfp,Romero-Shaw:2020owr}.
We report the Bayesian posterior distribution of the parameter $\nu$ in Fig.~\ref{post_nu}. The median value of the parameter, together with its $90\%$ equal-tailed credible interval, is $\nu = -1.1^{+3.9}_{-1.1}$, $\nu = 0.5^{+3.5}_{-2.6}$, and $\nu = 0.7^{+3.1}_{-2.3}$ for the Truncated, Power Law + Peak, and Broken Power Law mass models, respectively. Throughout this work, we report median values and $90\%$ credible intervals unless otherwise specified. The constraints on $\nu$ parameter are consistent across all three mass models, suggesting that our results are robust to the choice of mass distribution. However, the Power Law + Peak and Broken Power Law models yield slightly tighter constraints compared to the Truncated model. This difference can be attributed to the more complex structure of these mass models, which allows for a better fit to the observed BBH population. Compared to the constraint from the bright siren of GW170817, which yielded $-75.3 \leq \nu \leq 78.4$~\cite{Arai:2017hxj}, the constraint from the spectral siren method using GWTC-3 is improved by an order of magnitude. This improvement can be attributed to the higher redshifts of the BBH mergers in GWTC-3, with the farthest BBH reaching a redshift of $z \sim 1$~\cite{KAGRA:2021duu}, while the redshift of GW170817 is only $z \sim 0.008$~\cite{LIGOScientific:2017vwq}.

The improved constraint on $\nu$ has important implications for theories of modified gravity and cosmology. A value of $\nu$ consistent with zero, as suggested by our results, indicates that the propagation of GWs is not significantly modified compared to the predictions of GR. This can place constraints on alternative theories of gravity that predict a time-varying Planck mass or a running of the gravitational coupling strength \cite{Bellini:2014fua, Gleyzes:2014qga, Saltas:2014dha}. Such theories, which include scalar-tensor theories like Horndeski and beyond-Horndeski models, are motivated by the desire to explain the accelerated expansion of the Universe without invoking a cosmological constant \cite{Clifton:2011jh, Joyce:2014kja, Kase:2018aps, Lee:2022cyh}.

\begin{figure}[H]
	\centering
	\includegraphics[width=0.5\textwidth]{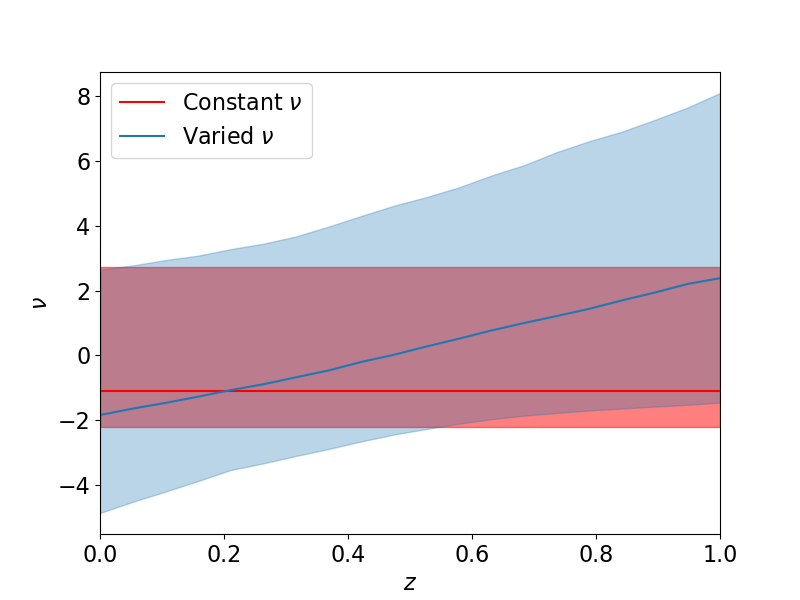}
	\caption{\label{ppd_nu}Posterior predictive distribution for $\nu(z)$ under the truncated mass model. The red shaded region shows the $90\%$ credible interval for the constant $\nu$ case, while the blue shaded region represents the redshift-dependent $\nu(z)$ parameterization given by Eq.~(\ref{nuz}). The solid curves indicate the median values for each model. The positive slope in the blue region suggests a mild increase in the friction parameter with redshift, though with large uncertainties due to the limited redshift range of GWTC-3.}
\end{figure}

Given that the redshifts of BBHs observed in GWTC-3 are relatively low ($z \lesssim 1$), we initially assumed a constant value of $\nu$ throughout the observable redshift range. However, our methodology can be readily extended to accommodate redshift-dependent scenarios. To demonstrate this capability, we consider a first-order Taylor expansion of $\nu(z)$ around $z = 0$:
\begin{equation}\label{nuz}
	\nu(z) = \nu_0 + \nu_1 z = \nu_0 - \nu_1 + \nu_1 (1+z),
\end{equation}
where $\nu_0$ represents the present-day value of the friction parameter and $\nu_1$ characterizes its evolution rate with redshift. As a demonstration of this extended framework, Fig.~\ref{ppd_nu} shows the posterior predictive distribution for $\nu(z)$ under the truncated mass model. The analysis yields constraints of $\nu_0 = -1.8^{+4.5}_{-3.0}$ and $\nu_1 = 4.2^{+5.2}_{-4.5}$, revealing a positive correlation between $\nu$ and redshift $z$. The positive value of $\nu_1$ suggests a mild increase in the friction parameter toward higher redshifts, though the large uncertainties reflect the limited constraining power of the current dataset. While this trend is intriguing, we caution that the relatively small redshift range and limited number of events in GWTC-3 restrict our ability to draw definitive conclusions about $\nu$ evolution. Future GW catalogs with larger event numbers and extended redshift coverage will enable more robust constraints on time-varying modified gravity effects.

Our results also have implications for the physics of the early Universe. A nonzero value of $\nu$ could affect the propagation of primordial GWs generated during inflation \cite{Boyle:2007zx, Nishizawa:2017nef, Nunes:2018zot}. The improved constraint on $\nu$ can be used to refine predictions for the spectrum of the stochastic GW background and its potential detectability by future GW observatories like LISA \cite{Bartolo:2016ami, Caprini:2018mtu, Caldwell:2018giq}.

The spectral siren method employed in this paper relies on the assumption that the BBH population follows a known redshift distribution, which is inferred from galaxy catalogs and cosmological simulations \cite{Fishbach:2018edt, Mukherjee:2020kki}. Uncertainties in this assumed redshift distribution can propagate to the constraint on $\nu$. In future work, it will be important to quantify the impact of these uncertainties and to explore alternative methods for inferring the redshift distribution of BBHs \cite{Mukherjee:2020hyn, Farr:2019twy, You:2020wju}.

Another potential source of systematic uncertainty is the calibration of the GW detectors. The strain calibration uncertainty can affect the measured luminosity distances and, consequently, the inferred constraint on $\nu$ \cite{Sun:2020wke, Vitale:2020gvb}. While the calibration uncertainty is sub-dominant compared to the statistical uncertainty for GWTC-3, it may become more significant as the number of detected events increases with future observing runs \cite{Purrer:2019jcp, Isi:2019asy}.

As the sensitivity of GW detectors continues to improve, the spectral siren method will become an increasingly powerful tool for constraining nonstandard GW propagation. The ongoing and upcoming observing runs of the LIGO-Virgo-KAGRA network (O4 and O5) are expected to yield a significant increase in the number of detected binary black hole mergers, reducing the statistical uncertainties on the $\nu$ parameter. Furthermore, third-generation ground-based detectors like the Einstein Telescope \cite{Punturo:2010zz} and Cosmic Explorer \cite{Reitze:2019iox} will be able to detect binary black hole mergers at even higher redshifts ($z>1$), enhancing the sensitivity to modified gravity effects. Space-based detectors such as LISA \cite{LISA:2017pwj}, Taiji \cite{Ruan:2018tsw} and TianQin \cite{TianQin:2015yph} will provide complementary constraints by probing the $\nu$ parameter through the detection of massive black hole binaries and extreme mass ratio inspirals~\cite{Liu:2023onj}. To fully harness the potential of these future datasets, it will be crucial to mitigate systematic uncertainties related to the assumed redshift distribution of BBHs and the calibration of GW detectors.

\section{Conclusion}\label{conclusion}
In conclusion, we have used the spectral siren method to constrain the parameter $\nu$, which characterizes the nonstandard propagation of GWs, using the GWTC-3 catalog of BBH mergers. Our results provide an order-of-magnitude improvement over previous constraints from the bright siren GW170817, placing limits on modified gravity theories and the physics of the early Universe. As the sensitivity of GW detectors improves and the number of detected events increases, the spectral siren method will continue to provide a powerful tool for testing GR and probing the nature of gravity on cosmological scales.

\Acknowledgements{%
We are very grateful to Shuxun Tian and Tao Zhu for the fruitful discussions that greatly improved the manuscript.
ZCC is supported by the National Natural Science Foundation of China under Grant No.~12405056, the Natural Science Foundation of Hunan Province under Grant No.~2025JJ40006, and the Innovative Research Group of Hunan Province under Grant No.~2024JJ1006. 
LL is supported by the National Natural Science Foundation of China (Grant No.~12505054 and ~12433001).}

\InterestConflict{The authors declare that they have no conflict of interest.}
\bibliographystyle{JHEP}
\bibliography{ref}
\end{multicols}
\end{document}